\documentclass[aps,showpacs,twocolumn,pre,superscriptaddress]{revtex4}
\usepackage{amsmath}
\usepackage{amssymb}
\usepackage{mathtext}
\usepackage{graphicx}
\usepackage{color}

\begin{document}
\newcommand {\e} {\varepsilon}
\newcommand {\sign} {\mbox{sign}}
\newcommand {\hb} {\hbar}
\newcommand {\Lit} {{\cal{L}}}
\renewcommand {\Re}{\mathrm{Re}}
\renewcommand {\Im}{\mathrm{Im}}

\title{Resonances and multistability in a Josephson junction connected to a resonator}

\author{Denis S.\ Goldobin}
\affiliation{Institute of Continuous Media Mechanics, UB RAS, Acad.\ Korolev Street 1, 614013 Perm, Russia}
\affiliation{Department of Theoretical Physics, Perm State University, Bukirev Street 15, 614990 Perm, Russia}
\author{Lyudmila S.\ Klimenko}
\affiliation{Institute of Continuous Media Mechanics, UB RAS, Acad.\ Korolev Street 1, 614013 Perm, Russia}
\affiliation{Department of Theoretical Physics, Perm State University, Bukirev Street 15, 614990 Perm, Russia}



\begin{abstract}
We study the dynamics of a Josephson junction connected to a dc current supply via a distributed parameter capacitor, which serves as a resonator. We reveal multistability in the current--voltage characteristic of the system; this multistability is related to resonances between the generated frequency and the resonator. The resonant pattern requires detailed consideration, in particular, because its basic features may resemble those of patterns reported in experiments with arrays of Josephson junctions demonstrating coherent stimulated emission. From the viewpoint of nonlinear dynamics, the resonances between a Josephson junction and a resonator are of interest because of specificity of the former; its oscillation frequency is directly governed by control parameters of the system and can vary in a wide range.  Our analytical results are in good agreement with the results of numerical simulations.
\end{abstract}


\maketitle

\thispagestyle{empty}

\section{Introduction}
Josephson junction---a contact between two superconductors optionally separated by a thin insulator layer---is a macroscopic element the dynamics of which is essentially quantum one~\cite{Josephson-1962,Zharkov-Al'tudov-1978}. These elements are natural voltage-to-frequency transducers. One can distinguish two regimes of operation of a Josephson junction (JJ): direct supercurrent with zero voltage applied across a junction and oscillations of supercurrent when the voltage is non-zero. The oscillation cyclic frequency for a dc voltage $V_\mathrm{dc}$ is $\omega=2eV_\mathrm{dc}/\hbar$, where $e$ is the electron charge and $\hbar$ is the Planck constant, and noteworthily indicates that charge carriers in superconductors are Cooper pairs.

From the viewpoint of nonlinear dynamics, JJ is a nonlinear oscillator with a very special property: its oscillation frequency in the ac mode is directly governed by a control parameter, the input current~\cite{Zharkov-Al'tudov-1978}. Though for the majority of nonlinear oscillators the frequency depends on the oscillation amplitude and control parameters, its variation is restricted to a certain range which rarely exceeds a few octaves. Hence, for the dynamics of a given nonlinear oscillator connected to a resonator, only a few or even none of resonant frequencies can be relevant.  In contrast, the oscillation frequency of a JJ oscillator varies nearly linearly with the input current, and any resonant frequencies are accessible and relevant as operation modes.
Thus, 
the dynamics of a single JJ connected to a resonator can be of generic interest.

Interconnected JJs were predicted to be able to self-synchronize with a common radiation field and emit coherently~\cite{Tilley-1970}. The suggested synchronization mechanism was a quantum one and analogous to the one in the case of superradiant atoms in resonant cavity. Even more similarity between these two quantum systems was revealed with further studies~\cite{Rogovin-Scully-1976,You-Nori-2011}.

Later on, observations of coherent emission for one- and two-dimensional arrays of junctions were reported~\cite{Jain-etal-1984,Benz-Burroughs-1991}, although the underlying synchronization mechanism was shown to be a classical one~\cite{Jain-etal-1984,Wiesenfeld-Benz-Booi-1994}.
Arrays of JJs turned out to be a remarkable object for the study of the classical synchronization and collective dynamics in populations of nonlinear oscillators. The reason for that is the property of populations of identical overdamped JJs which admit the employment of the Watanabe--Strogatz and Ott--Antonsen approaches~\cite{Watanabe-Strogatz-1994,Pikovsky-Rosenblum-2008,Marvel-Mirollo-Strogatz-2009,Ott-Antonsen-2008}.
These approaches allow deriving a low-dimensional self-contained system of ordinary differential equations for the order parameter and lend the opportunity for a significant advance in the study of generic laws of self-organization in collective dynamics on the basis of a rigorous mathematical treatment ({\it e.g.}, see~\cite{Marvel-Strogatz-2009,OKeeffe-Strogatz-2016,Dolmatova-Goldobin-Pikovsky-2017}).

\begin{figure}[b]
  \centerline{\includegraphics[width=0.44\textwidth]{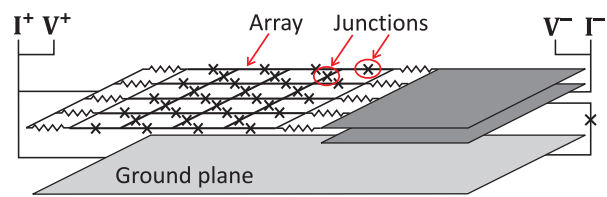}}
  \caption{Circuit of the array of Josephson junctions connected to an extensive capacitor
used in Ref.~\cite{Barbara-etal-1999} 
}
  \label{fig1}
\end{figure}

Observations for two- and one-dimensional arrays of JJs (see Fig.~\ref{fig1}), where stimulated emission was causing coherence, were presented in Refs.~\cite{Barbara-etal-1999,Vasilic-etal-2001,Vasilic-etal-2003}.
These observations could not be explained by classical coupling mechanisms and experimentally confirmed the predictions from Tilley~\cite{Tilley-1970} and Rogovin and Scully~\cite{Rogovin-Scully-1976}. The conclusion on the quantum nature of coherence in these experiments was firmly supported by the features of the current--voltage characteristics, the dependence of the emission power on the dc input power (also see~\cite{Vasilic-etal-2002}), and subtle analysis of the experimental set-ups.

It is interesting, that the electric circuits used in experiments~\cite{Barbara-etal-1999,Vasilic-etal-2001,Vasilic-etal-2002,Vasilic-etal-2003} contained a capacitor which could serve as a resonator with distributed parameters under certain conditions. All the components of the circuit were high-$Q$ elements.
In a high-$Q$ distributed parameter capacitor, the signals propagate with nearly no dispersion and decay. The interaction of self-sustained nonlinear oscillators with a neutrally stable dispersion-free waveguide is known to be able to lead to a rich and sophisticated resonant dynamics~\cite{Edelman-Gendelman-2013}.


The diversity in complex behavior of arrays of JJs creates a demand for a comprehensive picture of possible elementary collective and resonant phenomena in these arrays: macroscopic quantum coherence, classical synchronization, and resonances between JJ and high-$Q$ distributed-parameter elements. In this paper we consider the dynamics of a single JJ connected with a high-$Q$ resonator. As we will show below, the current--voltage characteristic of the latter system exhibits patterns with multiple resonances; some features of these patterns may distantly resemble those of the patterns reported for coherent states of junction array (cf.\ Fig.~\ref{fig3} of this paper and Fig.~2 in~\cite{Vasilic-etal-2003}). The detailed knowledge on the current--voltage characteristics of a single high-$Q$ JJ connected with a high-$Q$ resonator will complement the picture of elementary phenomena.

From the engineering point of view, the resonances in the system under our consideration are important for operation of JJ as a voltage-to-frequency or current-to-frequency transducer~\cite{Ozyuzer-etal-2007}; they can either affect susceptibility of the system to control or enhance the stability of generated frequencies.
Growing practical interest to JJs is also related to the construction of new tunable metamaterials~\cite{Butz-etal-2013,Jung-etal-2013,Pierro-Filatrella-2015}.


In this paper we derive the governing equations for the Josephson junction connected to a resonator (distributed parameter capacitor). Then the analytical solutions are obtained for the case of high generation frequency ({\it or} high input current) and low energy dissipation and confirmed with the results of numerical simulations. Further, we develop the weakly nonlinear analysis, which explains non-linear resonances observed with numerical simulation at low frequencies. Finally, we discuss the results and derive conclusions.

\section{Josephson junction with a~resonator}\label{sec2}
\subsection{Basic physical and mathematical model}
\begin{figure}[b]
  \centerline{\includegraphics[width=0.46\textwidth]{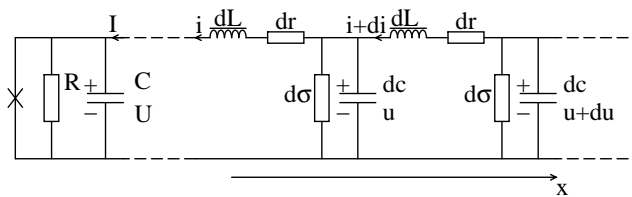}}
  \caption{Josephson junction
connected to a resonator with distributed capacity, induction, and ohmic resistance}
  \label{fig2}
\end{figure}

Let us consider an elongate resonator (distributed parameter capacitor) of length $l$ along the $x$-axis, connected to the Josephson junction at $x=0$ and an external current supply at $x=l$. The inductance $L$, the capacity $c$, the resistance $r$ for the current along the resonator, and the conductance $\sigma$ for the leakage current between its plates are distributed as shown in Fig.~\ref{fig2}. For the infinitesimal interval $dx$ the voltage increment $du$ and the current increment $di$ are
\begin{equation}
du=dL\,i_t+dr\,i\,,
\qquad
di=dc\,u_t+d\sigma\, u\,,
\label{eq1_1}
\end{equation}
where subscript $t$ indicates the partial time derivative. Hence,
\begin{equation}
i_t+r_xL_x^{-1}i=L_x^{-1}u_x\,,
\quad
u_t+\sigma_xc_x^{-1}u=c_x^{-1}i_x\,,
\label{eq1_2}
\end{equation}
where subscript $x$ indicates the partial $x$-derivative; $L_x$, $c_x$, $r_x$ and $\sigma_x$ are the inductance, capacity, resistance and leakage conductance per the unit length of the capacitor, respectively.

The net current $I$ through the Josephson junction is contributed by the tunnelling current $I_0\sin\phi$, the leakage current $U/R$, and the bias current $C(dU/dt)$ due to the junction electrical capacity~\cite{Zharkov-Al'tudov-1978};
\begin{equation}
I=I_0\sin{\phi}+\frac{U}{R}+CU_t\,,
\qquad
\phi_t=\frac{2e}{\hb}U\,,
\label{eq1_3}
\end{equation}
where $U$ is the potential drop on the junction, $\phi$ is the phase difference across the junction of the Ginzburg--Landau complex order parameter associated to the macroscopic current in a superconductor, $I_0$ is determined by physical properties of the junction, $R$ is the ohmic resistance of the junction, $C$ is the junction capacity, $e$ is the elementary charge, $\hb$ is the Planck constant.

It is convenient to make the following rescaling of coordinates and variables and introduce dimensionless parameters:
\begin{equation}\begin{array}{c}
\displaystyle
\displaystyle
 x=l\tilde{x}\,,\quad
 t=t_\ast\tilde{t}\,,\quad
 I=I_0\tilde{I}\,,\quad
 U=U_\ast\tilde{U}\,,\\[5pt]
\displaystyle
 t_\ast=\sqrt{\frac{\hb C}{2eI_0}}\,,\quad
 U_\ast=I_0\sqrt{\frac{L_x}{c_x}}\,,\quad
 v=\frac{t_\ast}{l\sqrt{c_xL_x}}\,,
\\[10pt]
\displaystyle
 \gamma_i=\frac{r_x t_\ast}{L_x},\
 \gamma_u=\frac{\sigma_x t_\ast}{c_x},\
 \beta=\frac{t_\ast}{2RC},\
 F
 =\sqrt{\frac{\hbar c_x}{2e C L_xI_0}}\,.
\end{array}
\label{eq1_4}
\end{equation}
Henceforth, the sign tilde is omitted.

In dimensionless form, Eqs.~(\ref{eq1_2}) and (\ref{eq1_3}) constitute the governing equation system with distributed parameters (one-dimensional) and boundary conditions:
\begin{eqnarray}
&u_t+\gamma_uu=vi_x\,,\label{ms_1}\\[5pt]
&i_t+\gamma_ii=vu_x\,,\label{ms_2}\\[5pt]
x=0:&i(0,t)=\sin{\phi}+2\beta\phi_t+\phi_{tt}\,,\label{ms_3}\\[5pt]
&\phi_t=F^{-1}u(0,t)\,,\label{ms_4}\\[5pt]
x=1:&i(1,t)=I_1(t)\,,\label{ms_5}
\end{eqnarray}
where $I_1(t)$ is the external input current.

\subsection{Waves in resonator}
We first focus on the case of $\gamma_i=\gamma_u=\gamma$. In this case we can seek for an analytical solution in form of a pair of counterpropagating decaying waves;
\begin{equation}
i(x,t)=e^{-\gamma t}(g(t-x/v)+h(t+x/v))\,.
\label{eq2_1}
\end{equation}
Eq.~(\ref{ms_5}) yields
\begin{equation}
h(t)=I_1(t-T)\,e^{\gamma(t-T)}-g(t-2T)\,,
\label{eq2_3}
\end{equation}
where $T=v^{-1}$. Substituting the latter equation into Eq.~(\ref{eq2_1}), one can find
\begin{eqnarray}
 i(x,t)=I_1(t-T+x/v)\,e^{-\gamma(T-x/v)}\qquad\qquad
\nonumber\\
 {}+e^{-\gamma t}\big(g(t-x/v)-g(t-2T+x/v)\big)\,.
\label{eq2_4}
\end{eqnarray}

One can seek for $u(x,t)$ in the same form as $i(x,t)$; specifically,
 $u(x,t)=e^{-\gamma t}\big(g_1(t-x/v)+h_1(t+x/v)\big)$.
From Eq.~(\ref{ms_1}) or Eq.~(\ref{ms_2}), $g_1'(\xi)=-g'(\xi)$ and $h_1'(\xi)=h'(\xi)$ (here the prime denotes derivative); therefore,
\begin{equation}
u(x,t)=e^{-\gamma t}\big(-g(t-x/v)+h(t+x/v)+const\big)\,,
\label{eq2_5}
\end{equation}
where $const$ can be set to zero by renormalization of $g$ and $h$. Substituting $h$, one obtains
\begin{eqnarray}
 u(x,t)=I_1(t-T+x/v)\,e^{-\gamma(T-x/v)}
\qquad\qquad
\nonumber\\
 {} +e^{-\gamma t}\big(-g(t-x/v)-g(t-2T+x/v)\big)\,.
\label{eq2_6}
\end{eqnarray}

For the general case of $\gamma_i\ne\gamma_u$ and a weakly dissipative resonator (which is of practical interest), {\it i.e.}, $\gamma_i\ll v$ and $\gamma_u\ll v$, Eqs.~(\ref{eq2_4}) and (\ref{eq2_6}) are still valid with
\[
\gamma=\frac{\gamma_i+\gamma_u}{2}
\]
up to corrections $\mathcal{O}\big((\gamma_i-\gamma_u)^2/v^2\big)$.

\subsection{Dynamics of Josephson junction}
Now we can recast the full set of the governing equations of our dynamic system, some of which are partial differential equations, into the form of an ordinary differential equation for phase $\phi(t)$ with time-delay terms. Using Eq.~(\ref{eq2_6}) one can rewrite Eqs.~(\ref{ms_3}) and (\ref{ms_4}) as
\begin{eqnarray}
 &&\hspace{-15pt}
 \phi_{tt}(t)+2\beta\phi_t(t)+\sin{\phi(t)}=I_1(t-T)e^{-\gamma T}
\nonumber\\
 &&\qquad\qquad\qquad
 {}+f(t)-e^{-2\gamma T}f(t-2T)\,,\label{eq3_1}\\[5pt]
 &&\hspace{-15pt}
 F\phi_t(t)=I_1(t-T)e^{-\gamma T}
\nonumber\\
 &&\qquad\qquad\qquad
 {} -f(t)-e^{-2\gamma T}f(t-2T)\,,\label{eq3_2}
\end{eqnarray}
where $f(t)=e^{-\gamma t}g(t)$. From Eq.~(\ref{eq3_2}),
\[
f(t)=-F\phi_t(t)+I_1(t-T)e^{-\gamma T}-e^{-2\gamma T}f(t-2T)\,.
\]
In this equation, one can substitute $f(t-2T)$ in the latter term with the expression for $f(t)$ taken for the time instant $t-2T$, and repeat this procedure for $t-4T$, $t-6T$, {\it etc.}, finally obtaining
$f(t)=\sum_{n=0}^\infty\left(-e^{-2\gamma T}\right)^n(-F\phi_t(t-2nT)+I_1(t-(2n+1)T)e^{-\gamma T})$. Substituting $f(t)$ into Eq.~(\ref{eq3_1}), one finds
\begin{eqnarray}
 \phi_{tt}(t)+(2\beta+F)\phi_t(t)+\sin{\phi(t)}\qquad\qquad\qquad
\nonumber\\[5pt]
 {} =2\sum\limits_{n=0}^{\infty}(-1)^ne^{-(2n+1)\gamma T}I_1(t-(2n+1)T)
\nonumber\\
 {} -2F\sum\limits_{n=1}^{\infty}(-1)^ne^{-2n\gamma T}\phi_{t}(t-2nT)\,.
\label{eq3_3}
\end{eqnarray}
For a constant in time input current $I_1(t)=I_1$ the first sum in Eq.~(\ref{eq3_3})
 $\sum_{n=0}^{\infty}(-1)^ne^{-(2n+1)\gamma T}=(2\cosh{\gamma T})^{-1}$;
thus one obtains
\begin{eqnarray}
\phi_{tt}(t)+(2\beta+F)\phi_{t}(t)+\sin{\phi(t)}
 =\frac{I_1}{\cosh{\gamma T}}\quad
\nonumber\\
 {}-2F\sum\limits_{n=1}^{\infty}(-1)^ne^{-2n\gamma T}\phi_{t}(t-2nT)\,.
\label{eq3_4}
\end{eqnarray}
Here and hereafter, we consider the case of constant input current $I_1$, which corresponds to typical experimental set-ups with dc current supplies. The chain of delayed terms of the form $\sum_{n=1}^\infty\rho^n x(t-n\tau)$ (where coefficient $|\rho|<1$, $x(t)$ is some system variable, and $\tau$ is the delay time) frequently appears for resonators, including interferometers, and is referred to as ``recursive delay feedback'' or ``extended delay feedback''.

The nonlinear differential equation (\ref{eq3_4}) with a linear recursive delay feedback governs the dynamics of the system we consider. Our further study is focused on solving this equation, examining properties of its solution, and their interpretation.

\paragraph*{Average (measured) input voltage.}
Let us consider the average value of the input voltage, which can be treated as a measured input voltage as oscillations about this value are high-frequency ones,
\begin{equation}
V_1=\langle{u(1)}\rangle
 =\langle{I_1(t)-2e^{-\gamma t}f(t-T)}\rangle\,,
\label{eq3_5}
\end{equation}
where $\langle\dots\rangle$ denotes averaging over time. From Eq.~(\ref{eq3_2}), one can find
\begin{eqnarray}
 -e^{-\gamma T}f(t-T)=\sum\limits_{n=1}^{\infty}\big[I_1(t-2nT)e^{-2n\gamma
T}(-1)^n
\nonumber\\
 {}
 -F\phi_t(t-(2n-1)T)e^{-(2n-1)\gamma T}(-1)^n\big]\,.
\label{eq3_6}
\end{eqnarray}
Since $\langle\phi_{t}\rangle$ is constant in time by definition, Eqs.~(\ref{eq3_5}) and (\ref{eq3_6}) yield
\begin{equation}
V_1=I_1\tanh{2\gamma T}
 +\frac{F\langle\phi_{t}\rangle}{\cosh{\gamma T}}\,.
\label{eq3_7}
\end{equation}

\section{The case of high input current}\label{sec3}

When the net current $I$ through the junction is large compared to the maximal tunnelling current $I_0$ [see Eq.~(\ref{eq1_3})], the ohmic contribution in the current is dominating. The nearly constant ohmic current yields a nearly constant voltage $U$ across the junction and, according to Eq.~(\ref{eq1_3}), phase $\phi(t)$ rotates quickly with some oscillations about the linear growth trend; one can seek for the solution in form
\begin{equation}
\phi(t)=\phi_0+\omega t+a\cos{\omega t}+\dots\,,
\label{eq4_1}
\end{equation}
assuming $I_1\gg 1$, $\omega\gg 1$ and $a\ll 1$, where the dots stand for higher-order harmonics, which are to be neglected. The term $\sin\omega t$ is removed by means of shifting the time offset; this shift is represented by constant $\phi_0$, which is yet to be found.

For calculation of $\sin\phi$ in Eq.~(\ref{eq3_4}), we employ the Jacobi--Anger expansion;
\begin{eqnarray}
&&
\cos(a\cos\omega t)=J_0(a)
 +2\sum_{n=1}^{\infty}(-1)^nJ_{2n}(a)\cos{2n\omega t}\,,
\nonumber\\
&&
\sin(a\cos\omega t)
 =2\sum_{n=0}^{\infty}(-1)^nJ_{2n+1}(a)\cos{(2n+1)\omega t}\,,
\nonumber
\end{eqnarray}
where $J_n(a)$ is the $n$-th order Bessel function of the first kind. Keeping only the constant-in-time term and the first harmonics in the Jacobi--Anger expansion, one finds
\begin{eqnarray}
\sin\phi=\sin(\phi_0+\omega t)\cos(a\cos\omega t)\qquad\qquad
\nonumber\\
{} +\cos(\phi_0+\omega t)\sin(a\cos\omega t)
\nonumber\\[5pt]
        =J_1(a)\cos\phi_0
        +J_0(a)\cos\phi_0\sin{\omega t}\qquad
\nonumber\\
 {}
        +J_0(a)\sin\phi_0\cos{\omega t}+\dots\,.
\end{eqnarray}
Then Eq.~(\ref{eq3_4}) reads
\begin{equation}
\begin{array}{l}
\displaystyle
-\omega^2a\cos{\omega t}
 +(2\beta+F)\omega(1-a\sin{\omega t})
\\[7pt]
\quad\displaystyle
     {} +J_1(a)\cos\phi_0
        +J_0(a)\cos\phi_0\sin{\omega t}
\\[3pt]
\qquad\displaystyle
     {} +J_0(a)\sin\phi_0\cos{\omega t}+\dots
=\frac{I_1}{\cosh{\gamma T}}
\\
\qquad\displaystyle
 {}-2F\sum\limits_{n=1}^{\infty}(-1)^ne^{-2n\gamma T}\omega
 \big(1-a\sin{\omega(t-2nT)}\big).
\end{array}\label{eq4_2}
\end{equation}

\begin{figure*}[!t]
  \quad
  {\sf (a)}\hspace{-10pt}\includegraphics[width=0.410\textwidth]{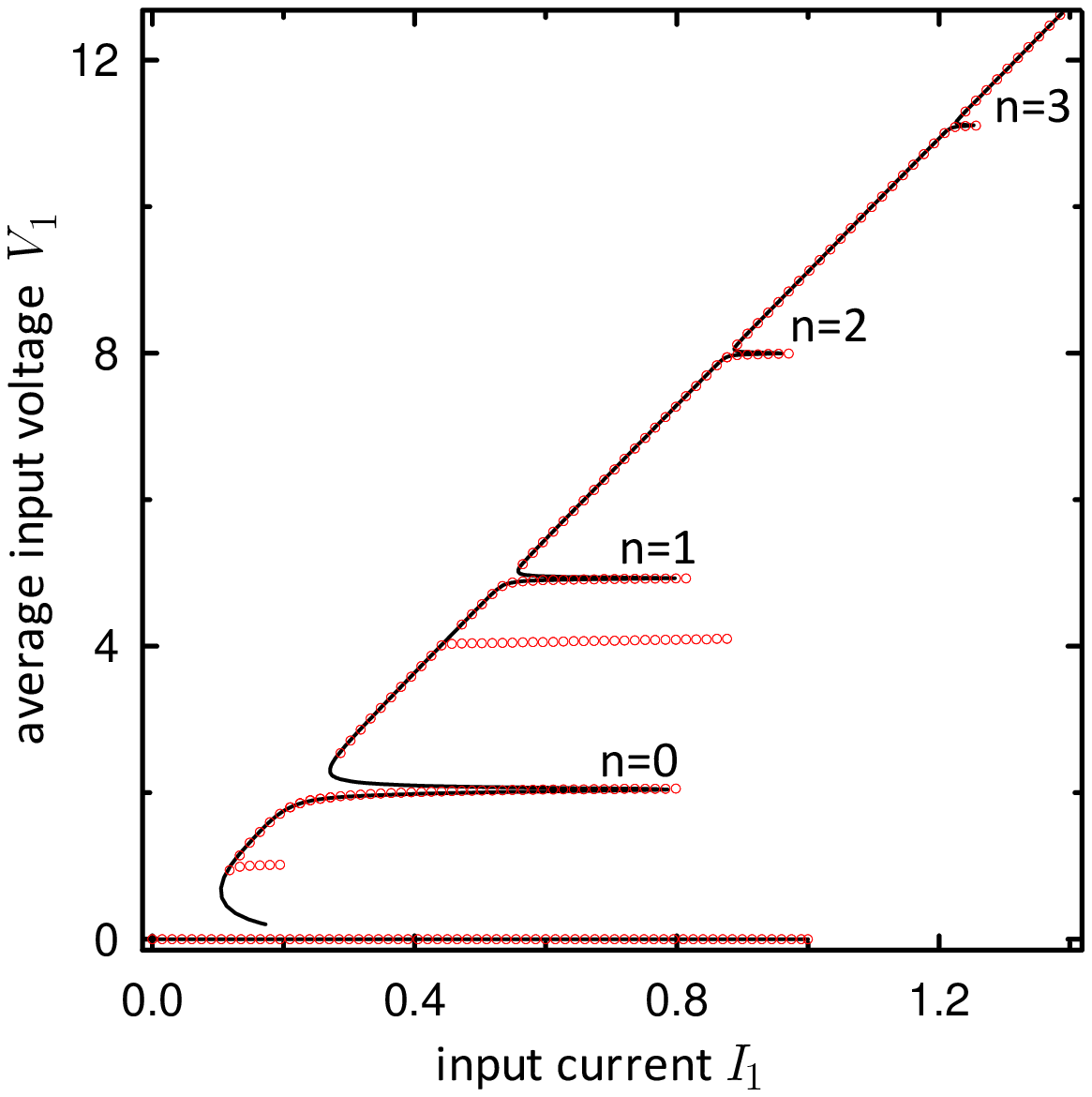}
  \qquad\qquad
  {\sf (b)}\hspace{-10pt}\includegraphics[width=0.412\textwidth]{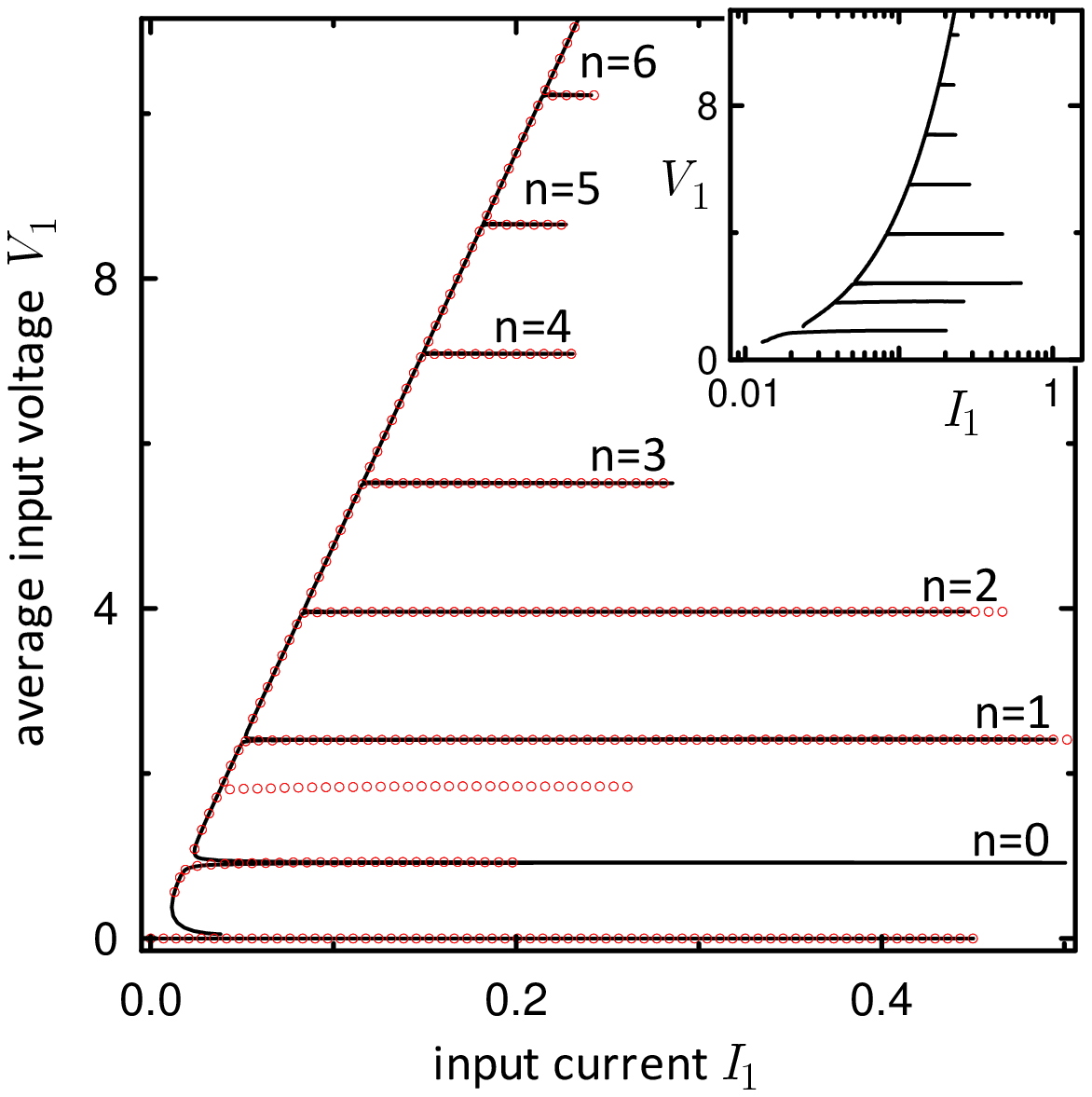}
  \caption{The current--voltage characteristic is plotted for (a) $\beta=0.05$, $\gamma=0.01$, $F=1$; (b) $\beta=0.005$, $\gamma=0.001$, $F=0.5$. The average input voltage $V_1$ is determined by Eq.~(\ref{eq3_7}). Red circles: the results of numerical simulation of Eq.~(\ref{eq3_4}), black solid line: the analytical solution (\ref{eq4_6})--(\ref{eq4_7}). In the insert graph the same current--voltage characteristic from numerical simulations is plotted with the log--linear scale to show the properties of peaks at nonlarge values of $V_1$. For non-large values of $V$, one can notice two small sharp stripes in numerical results deviating from the analytical solution; analytical description of these stripes requires the higher order corrections to be accounted for. With a recursive delay feedback, even weak anharmonicity is known to be able to lead to strong resonant effects~\cite{Goldobin-2011}. However, for moderate and large values of $V_1$ these high-order resonances are not detectable and the analytical theory describes the system dynamics well.  The dynamics of the Josephson junction in different resonant regimes is illustrated in Fig.~\ref{fig4}.
}
  \label{fig3}
\end{figure*}

Collecting constant-in-time terms and terms proportional to $\sin\omega t$ and $\cos\omega t$ in Eq.~(\ref{eq4_2}), to the leading order, one finds
\begin{eqnarray}
&\displaystyle
 (2\beta+F\tanh{\gamma T})\omega+J_1(a)\cos\phi_0
 =\frac{I_1}{\cosh{\gamma T}}\,,
\label{eq4_3}
\\[7pt]
&\displaystyle
 \omega\,\alpha_1(\omega)=\frac{J_0(a)}{a}\cos{\phi_0}\,,\quad
\label{eq4_4}\\[7pt]
&\displaystyle
 \omega\,\alpha_2(\omega)=\frac{J_0(a)}{a}\sin{\phi_0}\,,\quad
\label{eq4_5}
\end{eqnarray}
where
\[
\alpha_1(\omega)\equiv2\beta+\frac{F\sinh{2\gamma T}}{\cosh{2\gamma T}+\cos{2\omega T}}\,,
\]
\[
\alpha_2(\omega)\equiv\omega+\frac{F\sin{2\omega T}}{\cosh{2\gamma T}+\cos{2\omega T}}\,.
\]
One can recast Eqs.~(\ref{eq4_3})--(\ref{eq4_5}) in the form free from $\phi_0$;
\begin{eqnarray}
&\displaystyle\hspace{-10pt}
 (2\beta+F\tanh{\gamma T})\omega
 +\omega\,\alpha_1(\omega)\frac{a\,J_1(a)}{J_0(a)}
 =\frac{I_1}{\cosh{\gamma T}}\,,
\label{eq4_6}
\\[7pt]
&\displaystyle\hspace{-10pt}
 \frac{a}{J_0(a)}
 =\frac{1}{\omega\sqrt{\alpha_1^2(\omega)+\alpha_2^2(\omega)}}\,.\quad
\label{eq4_7}
\end{eqnarray}

For given value of $\omega$, Eq.~(\ref{eq4_7}) can be treated as a transcendental equation with respect to $a$. This equation possesses unique solution for $a$ within the range from $a=0$ to $2.4048...$, which is the first zero of the Bessel function $J_0(a)$. Since our derivations are valid for non-large $a$, we should restrict ourselves to the interior of the latter range. Thus, Eq.~(\ref{eq4_7}) dictates single-valued dependence of $a$ on $\omega$. With known $a(\omega)$, Eq.~(\ref{eq4_6}) yields the value of $I_1$ and Eq.~(\ref{eq3_7}) yields the value of $V_1$. Summarising, the high-frequency solution is parameterised by frequency $\omega$, which determines the amplitude $a$ of phase oscillation via transcendental equation (\ref{eq4_7}), and Eqs.~(\ref{eq4_6}) and (\ref{eq3_7}) yield values of the corresponding input current $I_1$ and the time-average input voltage $V_1$.

\subsection{The case of low energy dissipation in resonator}

Let us consider the case of $\gamma T\equiv\e\ll1$ in detail. In this case, one can simplify:
\[
\alpha_1=2\beta+\frac{F\e}{\e^2+\cos^2{\omega T}}\,,\quad
\alpha_2=\omega+\frac{F\tan{\omega T}}{1+\e^2\tan^2{\omega T}}\,,
\]

The expression $\alpha_2$ can turn to zero, which can result in resonantly high values of $I_1$. Let us find frequencies $\omega$, where $\alpha_2$ attains zero value. Condition $\alpha_2=0$ yields
\[
\omega+\e^2\omega\tan^2{\omega T}+F\tan{\omega T}=0\,,
\]
which can be viewed as a quadratic equation with respect to $\tan{\omega T}$. Hence, one can write
\[
\left(\tan{\omega T}\right)_{1,2}=\frac{-F\pm\sqrt{F^2-4\omega^2\e^2}}{2\e^2\omega}\,.
\]
For $\e\to0$, these two branches of roots take the limiting forms:
\begin{eqnarray}
&&\hspace{-10pt}
\tan{\omega_{1,n} T}=-\frac{\omega_{1,n}}{F}\,,
\label{eq4_9}
\\[7pt]
&&\hspace{-10pt}
\cot{\omega_{2,n} T}=-\frac{\e^2\omega_{2,n}}{F}\,.
\label{eq4_10}
\end{eqnarray}
The roots of these equations are
\[
\omega_{1,n}=\frac{\pi}{T}\left(n+\frac{1}{2}\right)+\frac{F}{\displaystyle \pi\left(n+\frac{1}{2}\right)}+...\,,
\]
\[
\omega_{2,n}=\frac{\pi}{T}\left(n+\frac{1}{2}\right)\left(1+\frac{\e^2}{FT}+...\right).
\]
For these roots, one finds
\[
\alpha_1(\omega_{1,n})\approx2\beta+\frac{\e}{F}(F^2+\omega_{1,n}^2)\,,
\quad
\alpha_1(\omega_{2,n})\approx2\beta+\frac{F}{\e}\,.
\]
At points where $\alpha_2=0$, Eq.~(\ref{eq4_6}) also simplifies to
\begin{equation}
I_1=(2\beta+\e F)\omega+J_1(a)\,.
\label{eq4_11}
\end{equation}

\begin{figure}[!t]
\centerline{
\includegraphics[width=0.40\textwidth]{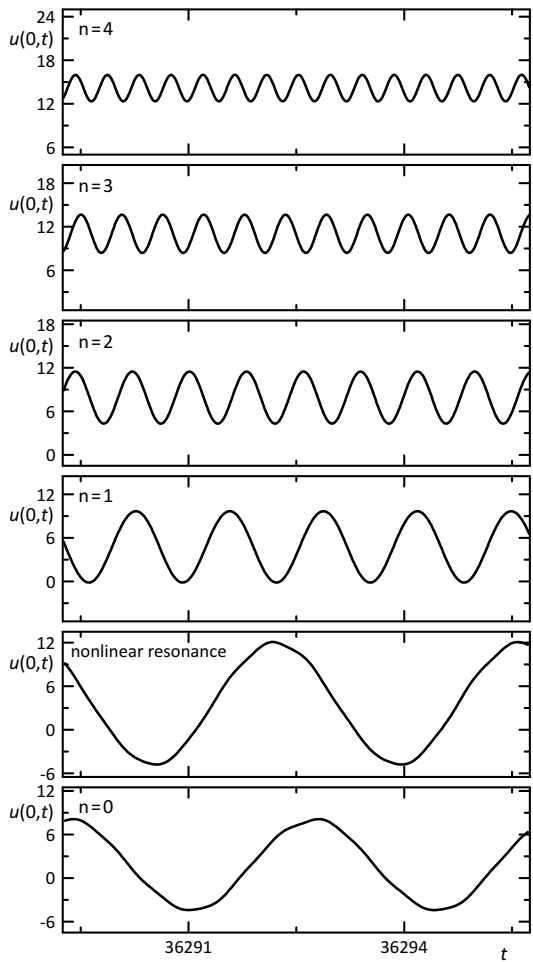}
}
  \caption{The dynamics of voltage across the Josephson junction $u(0,t)$ is simulated numerically for different resonant regimes, indicated with $n$, for input current $I_1=0.2$ and parameter values as in Fig.~\ref{fig3}b. The basic frequency of oscillations increases with $n$ as $\propto(n+1/2)$.}
  \label{fig4}
\end{figure}

One can see, that for the first group of roots, $\omega=\omega_{1,n}$, the value of $\alpha_1$ is small and Eq.~(\ref{eq4_7}) yields non-small values of $a$. Hence, $J_1(a)$ makes a non-small correction to the trend $(2\beta+\e F)\omega$.
Meanwhile, for the second group of roots, $\omega=\omega_{2,n}$, $\alpha_1$ is large and, according to Eq.~(\ref{eq4_7}), $a$ is small. Hence, $a(\omega_{2,n})\approx(\omega\alpha_1)^{-1}$ and
\[
I_1(\omega_{2,n})\approx
 (2\beta+\e F)\omega_{2,n}+\frac{\e}{2\omega_{2,n}F}\,.
\]
The increase of $I_1$ compared to the trend $(2\beta+\e F)\omega$ is small ($\propto\e$); there is no resonant peaks at $\omega_{2,n}$. Thus, there is a resonant increase of the input current $I_1$ at resonant frequencies $\omega=\omega_{1,n}$, this increase is especially strongly pronounced for small ohmic dissipation at the Josephson junction ($\beta\ll1$).

The physical mechanism of the increase of the input current required to maintain oscillations with resonant frequencies is as follows. With no dissipation and at resonant frequency, one can excite in the resonator a standing wave with zeros at the boundaries. For small dissipation and frequency mismatch, there are heirs of the resonant standing wave, which are the oscillating patterns with nearly zero values of fields at the boundaries. When one maintains not small, but moderate values of the fields at the boundaries (which are, in our case, due to inherent dynamics of the Josephson junction and external input current), the patterns in resonator are proportionally increased and become large-amplitude. Hence, even for small values of dissipation coefficients, the dissipation at the resonator becomes non-small and one requires stronger energy supply to the system to maintain the regime with a resonant frequency. This energy is supplied to the system with external input current, which has to be consequently increased.

\subsection{Comparison with numerical results and interpretation}

In Fig.~\ref{fig3}, one can see the results of the analytical theory [Eqs.~(\ref{eq3_7}), (\ref{eq4_6}), (\ref{eq4_7})] to match the results of numerical simulation well (the relative error of numerical simulations is below $10^{-12}$). The analytical theory inaccurately estimates the height of one or two low-frequency resonant peaks (while their position with respect to $V_1$ and, therefore, frequency are predicted accurately) and misses the nonlinear resonances which are non-negligible in the same low-frequency domain of parameters. The nonlinear corrections to the analytical theory are derived in the next section and with these corrections the nonlinear resonances appear where they are observed with numerical simulations. However, in the low-frequency domain, the series with respect to powers of $\omega^{-1}$ does not converge at the centres of peaks and the weakly-nonlinear analytical theory does not describe the system behavior; only the position of nonlinear resonances is predicted accurately.

It turns out that the analytical theory describes the resonant behavior very well immediately above the low-frequency domain (see Fig.~\ref{fig3}).

The analytical solution provides steady states, which can be either stable or unstable. At the solution branching points the tangential bifurcation occurs meaning the one of solutions is stable while the other is unstable. Since in numerical simulations, one observes only stable solutions, we can surely conclude that for resonant peaks the lower branch is stable, while the upper one is unstable (see Fig.~\ref{fig3}). A small distance between stable and unstable branches on the current--voltage plane does not mean that the attraction basin of the stable state is small; the branches are close only in the projection to this plane, while in the full phase space they are well remote from each other. With arbitrary initial conditions, the system frequently arrives to the stable resonant states.

\section{Nonlinear corrections of higher order}\label{sec4}

In this section we develop a perturbation analysis accounting for higher order terms. It will be convenient to read Eq.~(\ref{eq3_4}) in the form
\begin{equation}
\hat{L}\phi+\sin{\phi}=\frac{I_1}{\cosh{\gamma T}}\,,
\label{eq6_1}
\end{equation}
where
\begin{eqnarray}
\hat{L}\phi\equiv\phi_{tt}(t)+(2\beta+F)\phi_{t}(t)\qquad\qquad
\nonumber\\
 {}
 +2F\sum\limits_{n=1}^{\infty}(-e^{-2\gamma T})^n\phi_{t}(t-2nT)\,.
\nonumber
\end{eqnarray}

One can evaluate
\begin{equation}
\hat{L}\omega t=2\beta\omega+F\omega\tanh{\gamma T}\,,
\label{eq6_2}
\end{equation}
and
\begin{eqnarray}
&&\hspace{-20pt}
\hat{L}\phi_{\omega}=\omega\cos{\omega t}\bigg[-\omega a_{\omega} +2\beta b_{\omega}
\nonumber\\
&&\qquad
 {}
 +\frac{F(-a_{\omega}\sin{2\omega T}+b_{\omega}\sinh{2\gamma T})}{\cosh{2\gamma T}+\cos{2\omega T}}\bigg]
\nonumber\\
&&
 {}+\omega\sin{\omega t}\bigg[-\omega b_{\omega} -2\beta a_{\omega}
\nonumber\\
&&\qquad
 {}+\frac{F(-b_{\omega}\sin{2\omega T}-a_{\omega}\sinh{2\gamma T})}{\cosh{2\gamma T}+\cos{2\omega T}}\bigg].
\label{eq6_3}
\end{eqnarray}
where $\phi_{\omega}=a_{\omega}\cos{\omega t}+b_{\omega}\sin{\omega t}$.

After lengthy but straightforward calculations, one can find from Eq.~(\ref{eq6_1}), to the 3rd order,
\begin{eqnarray}
I_1=(2\beta\cosh{\gamma T}+F\sinh{\gamma T})\omega
\qquad\qquad\qquad
\nonumber\\[3pt]
 {}
 +\cosh{\gamma T}\Bigg(\frac{a_1^{(1)}+a_1^{(3)}}{2}
 +\frac{b_1^{(1)}a_2^{(2)}-b_2^{(2)}a_1^{(1)}}{4}
\nonumber\\
 {}
 -\frac{a_1^{(1)}\big[(a_1^{(1)})^2+(b_1^{(1)})^2\big]}{16}\Bigg)
 +\mathcal{O}(\omega^{-4})\,.\quad
\label{eq6_4}
\end{eqnarray}
Where $a_n^{(k)}$ and $b_n^{(k)}$ are determined by the following linear equations:
\begin{equation}
\left(\begin{array}{cr}
\alpha_2(\omega)&-\alpha_1(\omega)\\[5pt]
\alpha_1(\omega)&\alpha_2(\omega)
\end{array}\right)
\left(\begin{array}{c}
a_1^{(1)}\\[2pt]
b_1^{(1)}\end{array}\right)
=\frac{1}{\omega}\left(\begin{array}{c}
0\\[5pt]
1\end{array}\right)\,,
\label{eq6_5}
\end{equation}
\begin{equation}
\left(\begin{array}{cr}
\alpha_2(2\omega)&-\alpha_1(2\omega)\\[5pt]
\alpha_1(2\omega)&\alpha_2(2\omega)
\end{array}\right)
\left(\begin{array}{c}
a_2^{(2)}\\[2pt]
b_2^{(2)}\end{array}\right)
=\frac{1}{4\omega}\left(\begin{array}{c}
a_1^{(1)}\\[5pt]
b_1^{(1)}\end{array}\right)\,,
\label{eq6_6}
\end{equation}
\begin{eqnarray}
\left(\begin{array}{cr}
\alpha_2(\omega)&-\alpha_1(\omega)\\[5pt]
\alpha_1(\omega)&\alpha_2(\omega)
\end{array}\right)
\left(\begin{array}{c}
a_1^{(3)}\\[2pt]
b_1^{(3)}\end{array}\right)\qquad\qquad\qquad\qquad
\nonumber\\[5pt]
 {}
 =\frac{1}{\omega}\left(\begin{array}{c}
 a_2^{(2)}/2-a_1^{(1)}b_1^{(1)}/4\\[2pt]
 b_2^{(2)}/2-\big[(a_1^{(1)})^2+3(b_1^{(1)})^2\big]/8
\end{array}\right).
\label{eq6_7}
\end{eqnarray}

The average value of the input voltage is determined by Eq.~(\ref{eq3_7}) exactly;
\[
V_1=I_1\tanh{2\gamma T}+\frac{F\omega}{\cosh{\gamma T}}\,.
\]

Weakly-nonlinear solution (\ref{eq3_7}), (\ref{eq6_4})--(\ref{eq6_7}) provides corrections to the solution derived without accounting for $2\omega$- and higher harmonics. This solution is parameterised by frequency $\omega$. The weakly nonlinear solution correctly pinpoints the position of nonlinear resonances which can be seen in Fig.~\ref{fig3} (stripes without number $n$) for low frequencies which correspond to small average voltage $V_1$. Unfortunately, the weakly nonlinear solution helps only with identification of the position of nonlinear resonant peaks and confirming their nature; it does not reproduce the shape of peaks well, because of the divergence of the expansion with respect to $\omega^{-1}$ at low frequency domain.

\section{Conclusion}\label{concl}

A high--$Q$ circuit of a Josephson junction connected to resonator (a lengthy capacitor) has been found to exhibit multistability in regimes of operation and the current--voltage characteristic. The multistability is associated with tall peaks at the current--voltage characteristic emerging at generated oscillation frequencies which are resonant ones for a distributed parameter capacitor.

In resonant regimes, variation of the input current, which is a control parameter for this system in practice, makes a minor impact on the average input voltage and generation frequency. The resonant frequencies are given by Eq.~(\ref{eq4_9}), $\omega_{1,n}\approx(\pi/T)(n+1/2)$, and the corresponding average voltage determined by Eq.~(\ref{eq3_7}) reads $V_{1,n}\approx(\pi F/T)(n+1/2)$\,.

The detailed knowledge on features of the current--voltage characteristic we derived assists one to surely distinguish the resonant patterns we consider from the patterns reported for arrays of Josephson junctions in the lasing regimes of operation in~\cite{Barbara-etal-1999,Vasilic-etal-2001,Vasilic-etal-2003}. Currently, a thorough knowledge of the physical parameters of junctions is sufficient to identify the lasing regimes, as well as the dependence of the emission power on the dc input power for these regimes possesses recognizable properties. The information we report is most beneficial in the situations of the lack of quantitative information on the system parameters.

Considering Josephson junctions as natural voltage-to-frequency or current-to-frequency transducers, we would like to notice the possibility to strongly stabilize or efficiently control the generation frequency by means of a resonator. The stabilized generation frequencies are determined by generator properties; $\omega_{1,n}\approx(\pi/T)(n+1/2)$, where $T$ is the time of signal travel along the resonator.

\begin{acknowledgments}
The idea of considering this problem was suggested by Prof.\ Arkady Pikovsky, to whom the authors are also grateful for seminal discussions of the work findings and useful comments on the manuscript during the visit supported by G-RISC (grant No.\ M-2017a-3); the paper was also finalized during this visit.
The analytical derivations presented in Secs.~\ref{sec2}--\ref{sec3} and the numerical simulation have been performed under financial support by the Russian Science Foundation (Grant No.\ 14-21-00090).
\end{acknowledgments}







\end{document}